\documentclass[%
superscriptaddress, 
preprint,
amsmath,amssymb,
aps,
prb,
]{revtex4-2}

\usepackage{graphicx}
\usepackage{dcolumn}
\usepackage{bm}
\usepackage{gensymb}
\usepackage[mathlines]{lineno}

\begin{document}

\title{Epitaxial integration of improper ferroelectric h-YMnO$_3$ thin films in heterostructures}

\author{J. Nordlander}
\email{johanna.nordlander@mat.ethz.ch}
\affiliation{Department of Materials, ETH Zurich, CH-8093 Zurich, Switzerland}
\author{M. D. Rossell}
\author{M. Campanini}
\affiliation{Electron Microscopy Center, Empa, CH-8600 D\"{u}bendorf, Switzerland}
\author{M. Fiebig}
\author{M. Trassin}
\email{morgan.trassin@mat.ethz.ch}
\affiliation{Department of Materials, ETH Zurich, CH-8093 Zurich, Switzerland}

\begin{abstract}
We report on multiple fundamental qualitative improvements in the growth of improper ferroelectric hexagonal YMnO$_3$ (YMO) thin films and heterostructures by pulsed laser deposition (PLD). By a combination of pre-growth substrate annealing and low-energy-fluence PLD, we obtain a two-dimensional growth mode of the YMO films on yttria-stabilized zirconia (YSZ) with ultralow roughness and devoid of misoriented nanocrystalline inclusions. By inserting a sacrificial manganite layer capped with conducting indium-tin oxide between the substrate and the final film, the latter is grown in a fully lattice-relaxed mode and, thus, without any misfit dislocations while maintaining the extraordinary flatness of the films grown directly on pre-annealed YSZ. This provides a template for the fabrication of heterostructures based on hexagonal manganites as promising class of multiferroics with improper room-temperature ferroelectricity and the implementation of these into technologically relevant epitaxial metal\,$|$\,ferroelectric-type multilayers.

\end{abstract}

\maketitle

\clearpage

\section{Introduction}

Materials possessing coexisting ferroelectric and magnetic order are interesting for their rich physics and the great potential for novel technological applications resulting from magnetoelectric cross-coupling effects. The hexagonal rare-earth manganites, RMnO$_3$ (RMO) with R = Sc, Y, In, Dy-Lu, are textbook examples of such so-called multiferroic compounds. Here, improper ferroelectricity emerges between 1250 and 1650\,K \cite{Lilienblum2015,Chae2012} through its coupling to a non-ferroelectric lattice trimerization. A coexistence of this improper ferroelectric state with antiferromagnetic Mn$^{3+}$ order is seen below temperatures between 65 and 130\,K \cite{Lorenz2013}. The hexagonal manganites have been studied as bulk crystals and are particularly famous for their characteristic, topologically protected six-fold vortex trimerization--polarization domain pattern \cite{Choi2010,Jungk2010}, unconventional domain-wall conductance \cite{Meier2012a} and coupling of ferroelectric and antiferromagnetic domains \cite{Fiebig2002}. Hence, these materials can be the source of exotic ferroelectric functionalities not found in conventional ferroelectrics. Exploiting such phenomena in applications, however, depends on the implementability of RMO as single-crystal thin film in epitaxial heterostructures. In the technologically relevant ultrathin regime, it was shown that the epitaxial constraints imposed by the substrate on YMO films can significantly alter the structural distortions related to the primary-order lattice trimerization, which in turn affects the emergence of the improper electric polarization \cite{Nordlander2019}. The main challenge in studying such nanoscale effects and putting these films to use in devices resides in the difficulty to find lattice-matching substrates and suitable electrode materials promoting high-quality epitaxy of the hexagonal phase. So far, epitaxial hexagonal RMO films have been realized on substrates including (111)-oriented yttria-stabilized zirconia (YSZ), Si, Pt, MgO and $c$-cut Al$_2$O$_3$ \cite{Fujimura1996a,Dho2004,Marti2007,Imada1998,Cheng2018,Imada2001,Wu2011,Dubourdieu2007,Laukhin2006}. However, in many of these cases, the films exhibited a tendency to form secondary crystalline phases or orientations. Because the hexagonal manganites are uniaxial ferroelectrics with a polarization along the $c$-axis, strictly $c$-axis-oriented films are required, in addition to precise control of layer thickness at the nanoscale, for optimal implementation in heterostructures for oxide-electronics applications.

Here, choosing YMO as model compound, we demonstrate a route towards layer-by-layer growth of $c$-oriented, single-crystal ultrathin hexagonal manganite films on (111)-oriented YSZ, both with and without use of a conducting indium-tin oxide (ITO) buffer layer.

\section{Experimental details}

To grow our YMO films, we use pulsed laser deposition (PLD). The growth mode is monitored in-situ using reflection high-energy electron diffraction (RHEED). The topography of the films is measured using atomic force microscopy (AFM) in tapping mode (Bruker MultiMode 8). The crystalline structure of the films is characterized by x-ray diffraction (XRD) using a Panalytical X’Pert$^3$ MRD four-circle diffractometer and by high-angle annular dark-field scanning transmission electron microscopy (HAADF-STEM). Samples for STEM analysis are prepared by means of a FEI Helios NanoLab 600i focused-ion beam (FIB) instrument operated at accelerating voltages of 30 and 5 kV. A FEI Titan Themis microscope with a probe CEOS DCOR spherical aberration corrector operated at 300 kV is used for atomic-resolution HAADF-STEM data acquisition. A probe semiconvergence angle of 18 mrad is used in combination with an annular semidetection range of the annular dark-field detector set to collect electrons scattered between 66 and 200 mrad. Geometric phase analysis of the HAADF-STEM images is performed using the GPA tool contained in the FRWRtools plugin \cite{FRWRtools} for Gatan Digital Micrograph. The g(004)YMO and g(030)YMO peaks are used for analysis with a reference lattice set in the YSZ substrate and the following parameters: resolution = 1.2\,nm, smoothing = 5.0. We finally probe the improper ferroelectric polarization of the YMO films using optical second harmonic generation (SHG). A pulsed Ti:Sapphire laser at 800\,nm with a pulse duration of 45\,fs and repetition rate of 1\,kHz is converted using an optical parametric amplifier to a probe wavelength of 860\,nm. This probe beam is incident on the sample with a pulse energy of 20\,$\mu$J on a spot size 250\,$\mu$m in diameter. The generated light intensity is subsequently detected using a monochromator set to 430\,nm and a photomultiplier system.

\section{Results}

\begin{figure}
    \centering
    \includegraphics[scale=0.48]{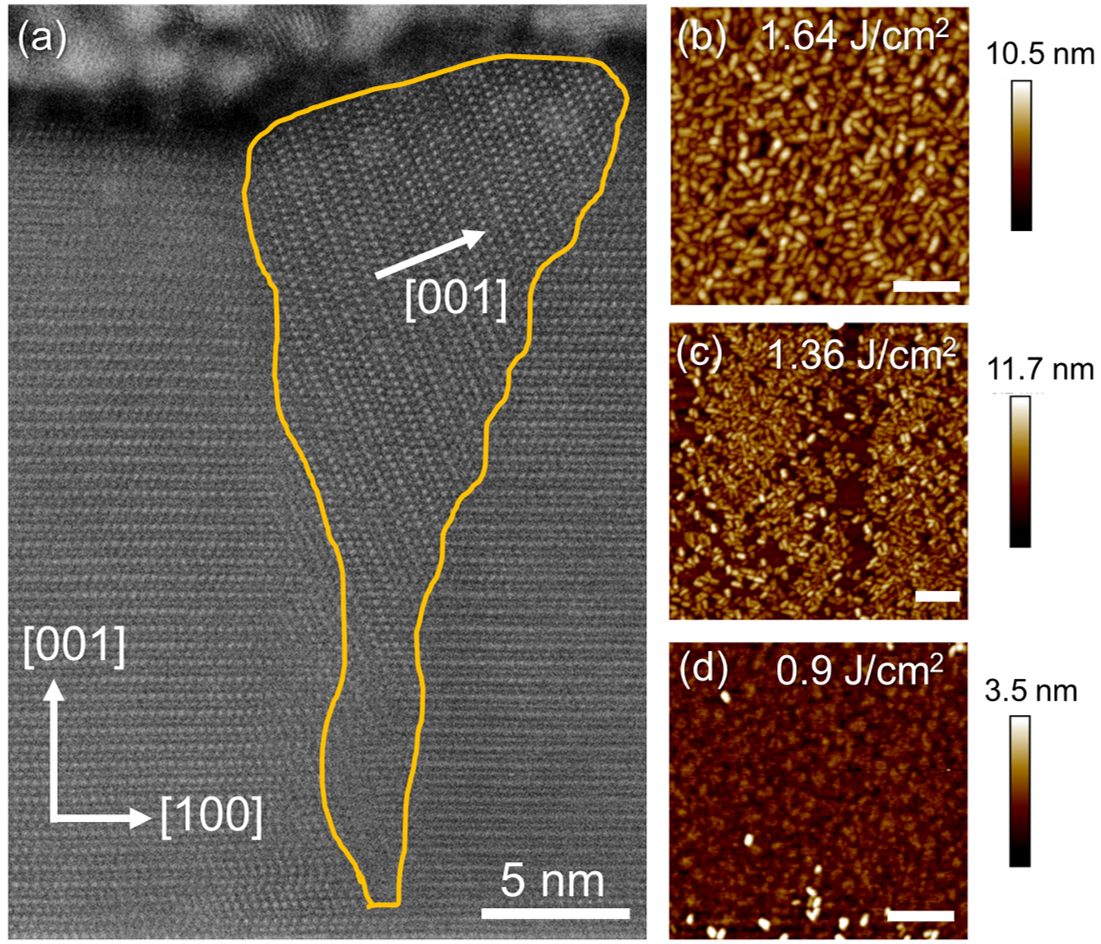}
    \caption{(a) HAADF-STEM image of a hexagonal YMO thin film showing the presence of a misoriented grain, outlined in yellow. (b-d) AFM topography scans of YMO films that were grown on as-received YSZ substrates at different laser fluences. The scale bar is 200 nm.}
    \label{fig:PLDfluence}
\end{figure}

We first optimize the YMO growth parameters on as-received, commercial (111)-oriented YSZ substrates (CrysTec GmbH). We use a stoichiometric ceramic target for laser ablation where the substrate is kept at 750\degree C in an oxygen partial pressure of 0.1 mbar. While the (111)-oriented cubic lattice has a hexagonal surface symmetry, perfect matching with the hexagonal lattice of the manganites remains challenging. For example, the presence of nanoinclusions with secondary $c$-axis orientations due to multiple options for lattice-matching on YSZ substrates have been reported \cite{Kordel2009,Jehanathan2010}. Such nanoinclusions are also observed in some of the earlier films in this work, see Fig.~\ref{fig:PLDfluence}(a). We thus start our investigation by tracking their density as function of the deposition conditions using AFM. The surface topography of YMO films grown at different laser fluences is shown in Fig.~\ref{fig:PLDfluence}(b-d). We find that a high laser fluence results in a rough surface with a high density of misoriented nanocrystals. As the fluence is decreased, less nanocrystals are observed and below 1\,J/cm$^2$ per pulse they are hardly detected anymore. In this state, we also find a smooth surface with a roughness of 0.34 nm, which therefore is likely to be correlated to the reduction in the density of misoriented grains inside the films.

\begin{figure}
    \centering
    \includegraphics[scale=0.47]{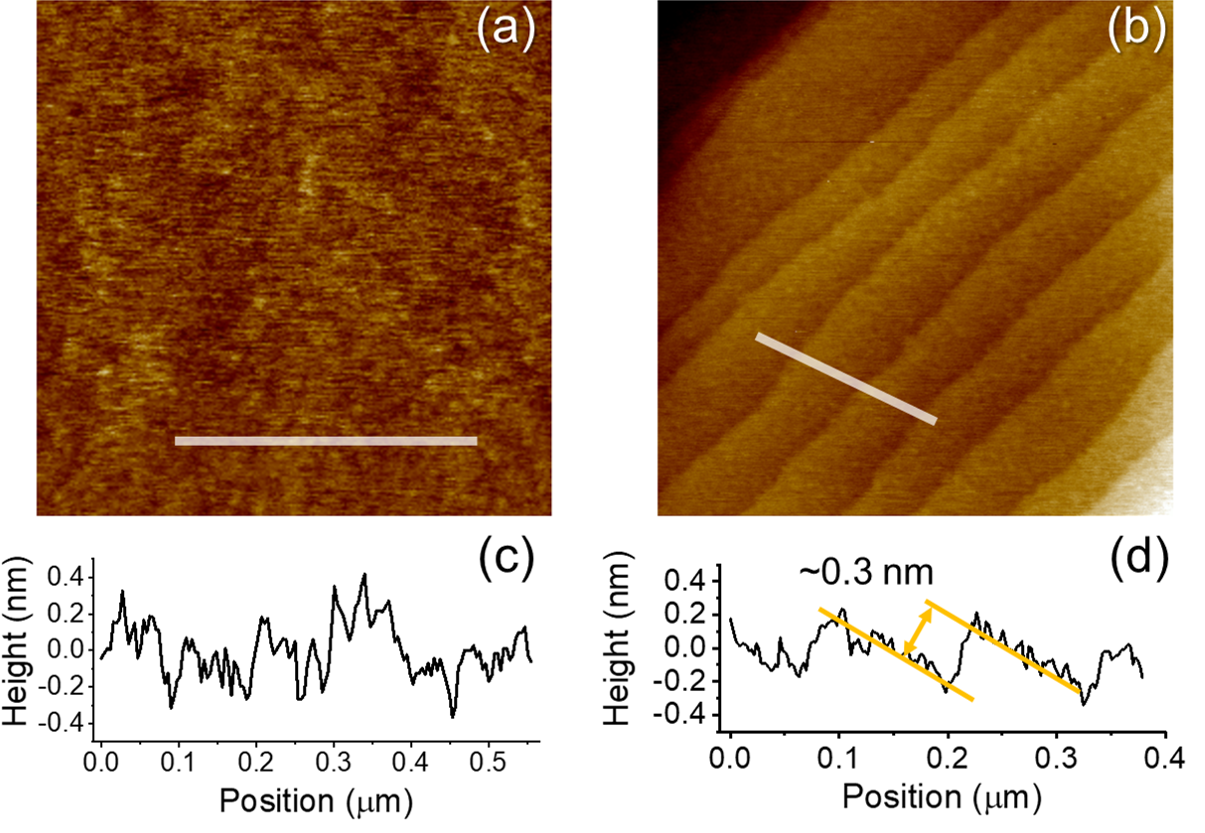}
    \caption{AFM topography scans of an YSZ substrate in as-received condition (a) before and (b) after subjecting it to thermal annealing. The scanned area is 1 $\mu$m$^2$. Line scans at the locations of the white lines are shown in (c) and (d), respectively. After annealing, the substrate exhibits a strikingly improved surface quality with steps of a height of about 0.3 nm, corresponding the the distance between the (111) crystallographic planes.}
    \label{fig:annealing}
\end{figure}

In order to improve the smoothness and single-crystallinity of the films further, we investigate the influence of the substrate surface topography on the YMO film growth. While the as-received YSZ substrates already have a low roughness of 0.16 nm, further improvement of the surface quality may facilitate island nucleation for a layer-by-layer growth mode \cite{Ohring1992}. We therefore induce surface reconstruction by annealing the substrates in air at 1250\degree C for 12 hours \cite{Honke2004}. As seen in Fig.~\ref{fig:annealing}, this annealing step significantly improves the surface morphology of the YSZ substrates. Specifically, it results in the formation of terraces with the step height of about 0.3 nm, corresponding to the distance between the (111) lattice planes in YSZ. 

\begin{figure}
    \centering
    \includegraphics[scale=0.57]{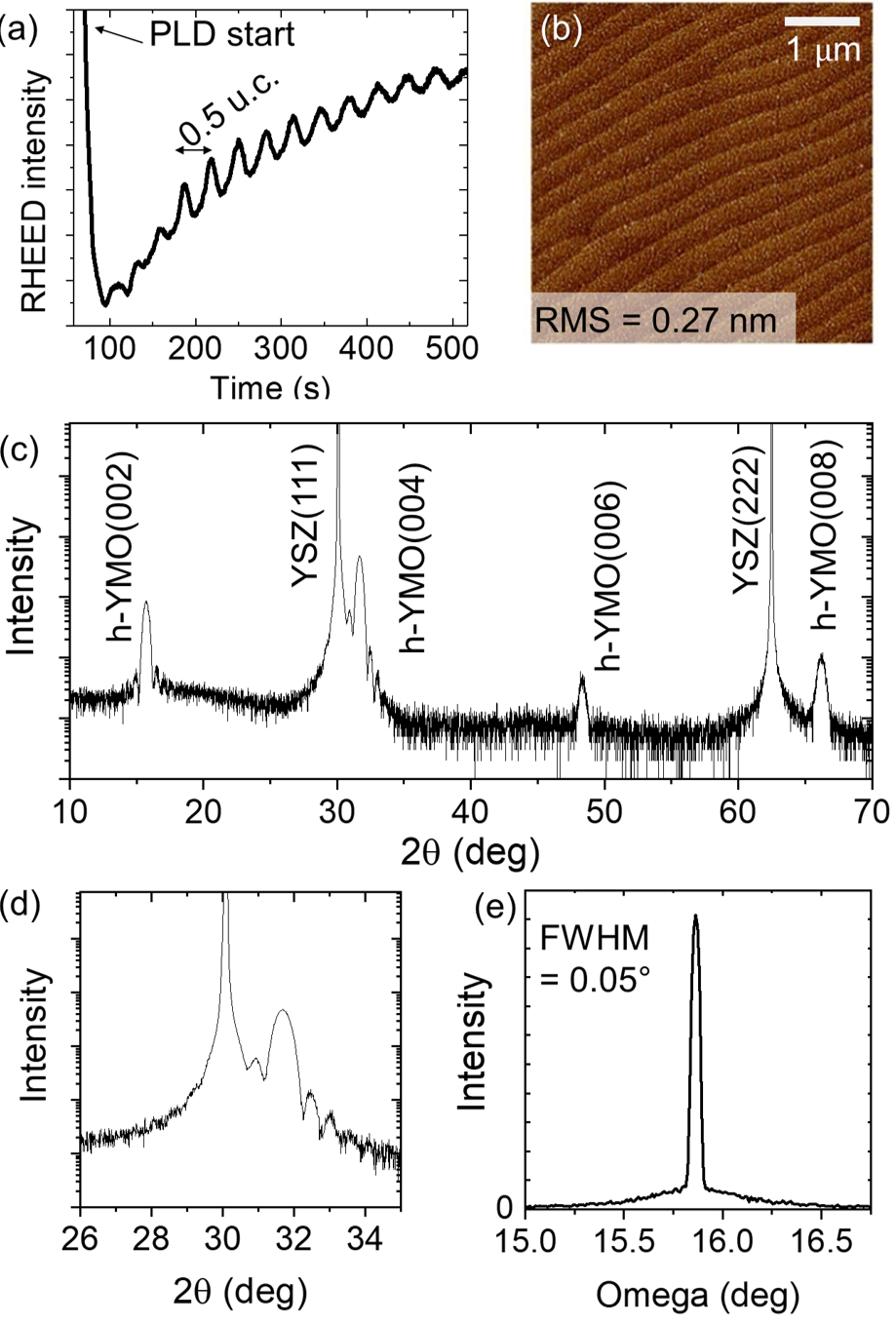}
    \caption{(a) RHEED intensity during YMO growth. The layer-by-layer-type growth mode is seen by the intensity oscillations, where each period corresponds to half a unit cell. (b) AFM topography scan of an YMO film with a thickness of 15 unit cells. (c) $\theta/2\theta$ XRD scan showing hexagonal (00$l$) peaks only. (d) High-resolution $\theta/2\theta$ scan around YSZ(111) and YMO(004). Laue oscillations around the film peak confirm the exceptional smoothness of the interfaces. (e) Rocking curve of YMO(004) with a narrow FWHM of 0.05\degree, further indicating a highly oriented YMO film. The corresponding FWHM of YSZ(111) is 0.009\degree.}
    \label{fig:YMO_YSZ}
\end{figure}

Applying the same growth protocol as with the as-received substrates, we use RHEED monitoring to asses the growth mode for the deposition of YMO on the annealed YSZ substrates. The RHEED intensity oscillations in Fig.~\ref{fig:YMO_YSZ}(a) indicate a layer-by-layer type growth mode. Film-thickness calibration by post-deposition x-ray reflectivity measurements reveals that each deposited monolayer is half a unit cell in height. Hence, in this growth mode, we achieve thickness control with sub-unit-cell precision. Most strikingly, the surface of the YMO films grown on the annealed YSZ substrates preserves the step-like morphology of the substrate and thus exhibits an ultralow roughness of less than 0.3\,nm (Fig.~\ref{fig:YMO_YSZ}(b)). Furthermore, x-ray $\theta/2\theta$ scans reveal a $c$-oriented YMO film with no trace of secondary phases or orientations (Fig.~\ref{fig:YMO_YSZ}(c)). The high quality of the film is further confirmed through XRD characterization. The visibility of thickness (Laue) oscillations around the YMO film peaks indicates sharp interfaces and the narrow rocking curve with a full-width at half-maximum (FWHM) of 0.05\degree\ points to a low mosaicity, see Fig.~\ref{fig:YMO_YSZ}(d,e).

\begin{figure}
    \centering
    \includegraphics[scale=0.47]{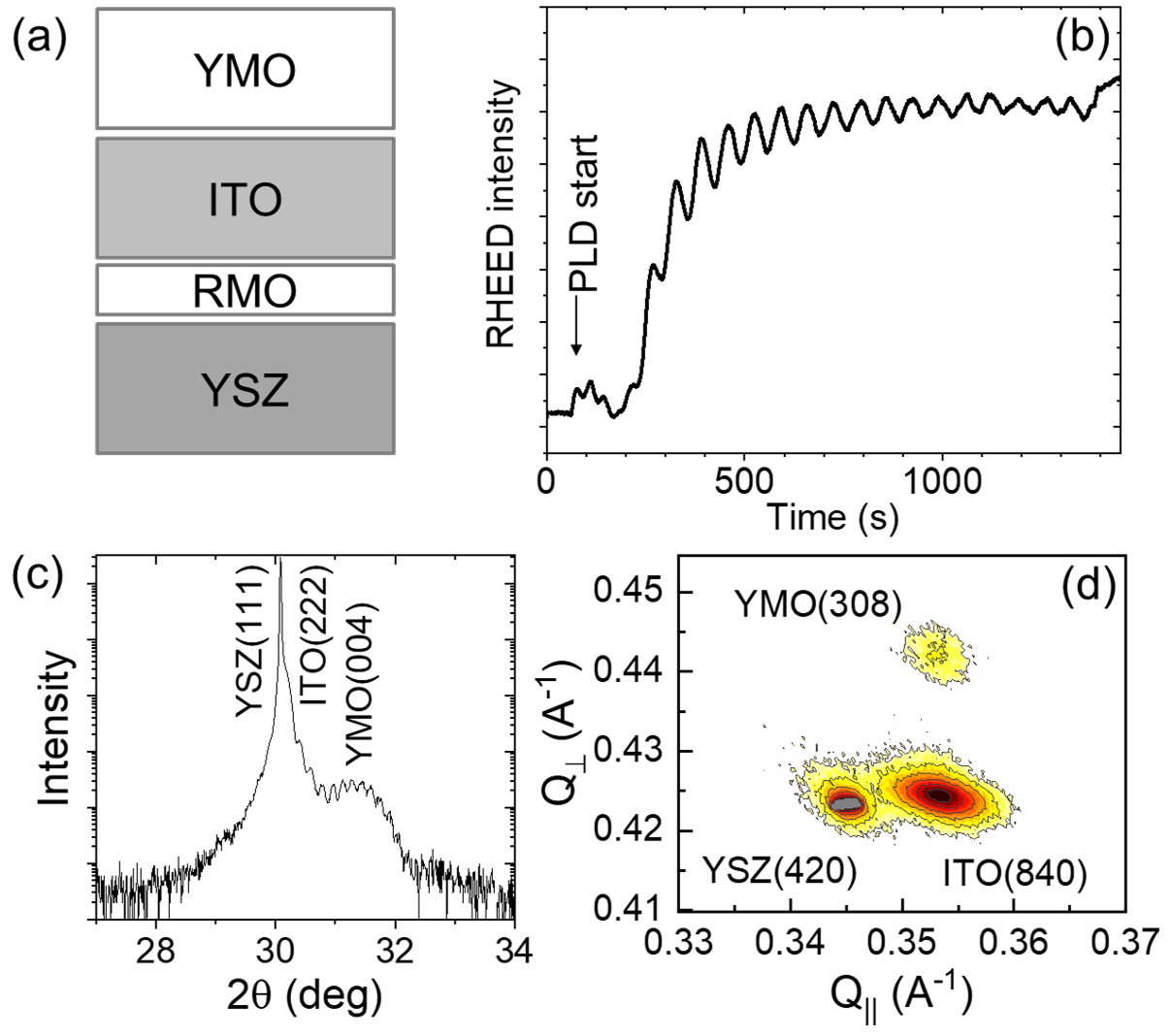}
    \caption{(a) Schematic of the YMO\,$|$\,ITO\,$|$\,RMO\,$||$\,YSZ heterostructure with RMO as sacrificial layer. (b) RHEED oscillations during YMO growth on the ITO bottom electrode indicate a layer-by-layer growth mode. (c) High-resolution $\theta/2\theta$ scan reveals Laue oscillations around the ITO(222) reflection. (d) Reciprocal space mapping of the YMO(308) reflection in the vicinity of the YSZ and ITO peaks. ITO and YMO adopt the same in-plane lattice parameter, yet relaxed against the YSZ substrate.}
    \label{fig:bottomElectrode}
\end{figure}

Having optimized the growth of ultrathin YMO films directly on YSZ, we now move on to identify a bottom-electrode material that allows us to preserve the excellent epitaxy and the layer-by-layer growth mode of the YMO. Here we propose Sn-doped In$_2$O$_3$, ITO (using In$_2$O$_3$ doped with 10 wt\% SnO$_2$ as target), as the material of choice. As seen in Table~\ref{tab:table1},\begin{table}[b]
\caption{\label{tab:table1}%
Bulk lattice parameters of the substrates (S) with relation to the YMO lattice for (001)$_\text{YMO}||$(111)$_\text{S}$ and $<\overline{\text{1}}\overline{\text{2}}0>_\text{YMO}||<1\overline{\text{1}}0>_\text{S}$. The equivalent hexagonal lattice parameters of YMO for both substrates are given as well as the corresponding lattice mismatch.
}
\begin{ruledtabular}
\begin{tabular}{lllc}
\textrm{Param.}&
\textrm{Cubic $a$ (\AA)}&
\textrm{Hex. $a$ (\AA)}&
\textrm{Lattice mismatch (\%)}\\
\colrule
YMO & - & 6.14 & - \\
YSZ & 5.12 & 6.27 & $+$2.1 \\
In$_2$O$_3$ & 10.117 & 6.20 & $+$0.98 \\
\end{tabular}
\end{ruledtabular}
\end{table} the bulk lattice constant of ITO, here taken as that of In$_2$O$_3$, lies close to that of YMO, with a lattice mismatch of less than 1\%, suggesting enhanced compatibility with the YMO lattice. As described in Ref. \onlinecite{Nordlander2019}, the YMO films grown directly on YSZ are not strained to the substrate, but rather adopt bulk-like in-plane lattice parameters by forming misfit dislocations at the YSZ interface. Remarkably, despite these local lattice defects, our YMO films maintain a smooth surface. This opens up a strategy towards obtaining YMO films of a fundamentally improved epitaxial quality. The strategy is as follows. We first grow an ultrathin rare-earth hexagonal manganite ``sacrificial layer" on top of the YSZ. The purpose of this RMO layer, which is not limited to R=Y, is to accommodate to the intrinsic lattice constants of the hexagonal manganites through the development of misfit dislocations but keep the surface flatness of the substrate. On top of this sacrificial RMO layer, we deposit the ITO with a threefold purpose. First, with its small lattice mismatch to the RMO compounds, it can adapt to the lattice constant and the surface flatness of the RMO layer below. Second, it poses a barrier against propagation of the misfit dislocations from the sacrificial RMO into the final YMO layer grown on top. Third, with its high conductivity, it acts as bottom electrode of an YMO heterostructure. With this strategy, the final YMO layer, deposited on top of ITO, will grow with full lattice relaxation and, most importantly, without misfit dislocations. We will hence complete our YMO\,$|$\,ITO\,$|$\,RMO\,$||$\,YSZ heterostructure [see Fig.~\ref{fig:bottomElectrode}(a)] with an YMO layer of expected drastically improved quality. At the same time, we note that the option to choose the RMO compound acting as sacrificial layer from the series of hexagonal manganites with slightly differing in-plane lattice constants \cite{Lorenz2013} further presents an opportunity to introduce a controlled, moderate lattice strain into the final YMO layer.

For verifying the proposed growth strategy, we grow the ITO films at a substrate temperature of 800\degree C and at a laser energy fluence of 0.7\,J/cm$^2$ per pulse with an oxygen partial pressure of 0.12 mbar. We find that ITO films grown on top of an ultrathin  ($\lesssim$ 2 unit cells) RMO buffer layer on the YSZ(111) substrate indeed adopt to the lattice constant of the RMO buffer layer (here, equivalent results were obtained for R = Y, Tb, In). In this case, no strain-relaxation is seen with increasing ITO thickness up to 50\,nm and ultraflat, (111)-oriented ITO films were obtained, with the steps in the topography corresponding to the $\sim$0.3\,nm distance between the (222) crystallographic planes of ITO. We note that the route presented above to achieve ultraflat ITO films that are lattice matched to the hexagonal manganites may also find its application in achieving improved surface quality of epitaxial ITO-based transparent electrodes in general \cite{Ohta2000,Kim2002}.

As next step, we find that our YMO films grown on top of this ITO conducting layer, still at 800\degree C and an oxygen partial pressure of 0.12 mbar, retain a similar layer-by-layer growth mode as on YSZ [Fig.~\ref{fig:bottomElectrode}(b)]. Thus, growth of highly oriented hexagonal YMO with smooth interfaces is obtained also in the case of an RMO-buffered ITO film as growth template. Indeed, x-ray reciprocal space mapping around the YMO(308) reflection confirms the epitaxial relationship in the trilayer system. We find that the thin-film layers all have matching in-plane lattices, yet relaxed against the YSZ substrate.

\begin{figure}
    \centering
    \includegraphics[scale=0.4]{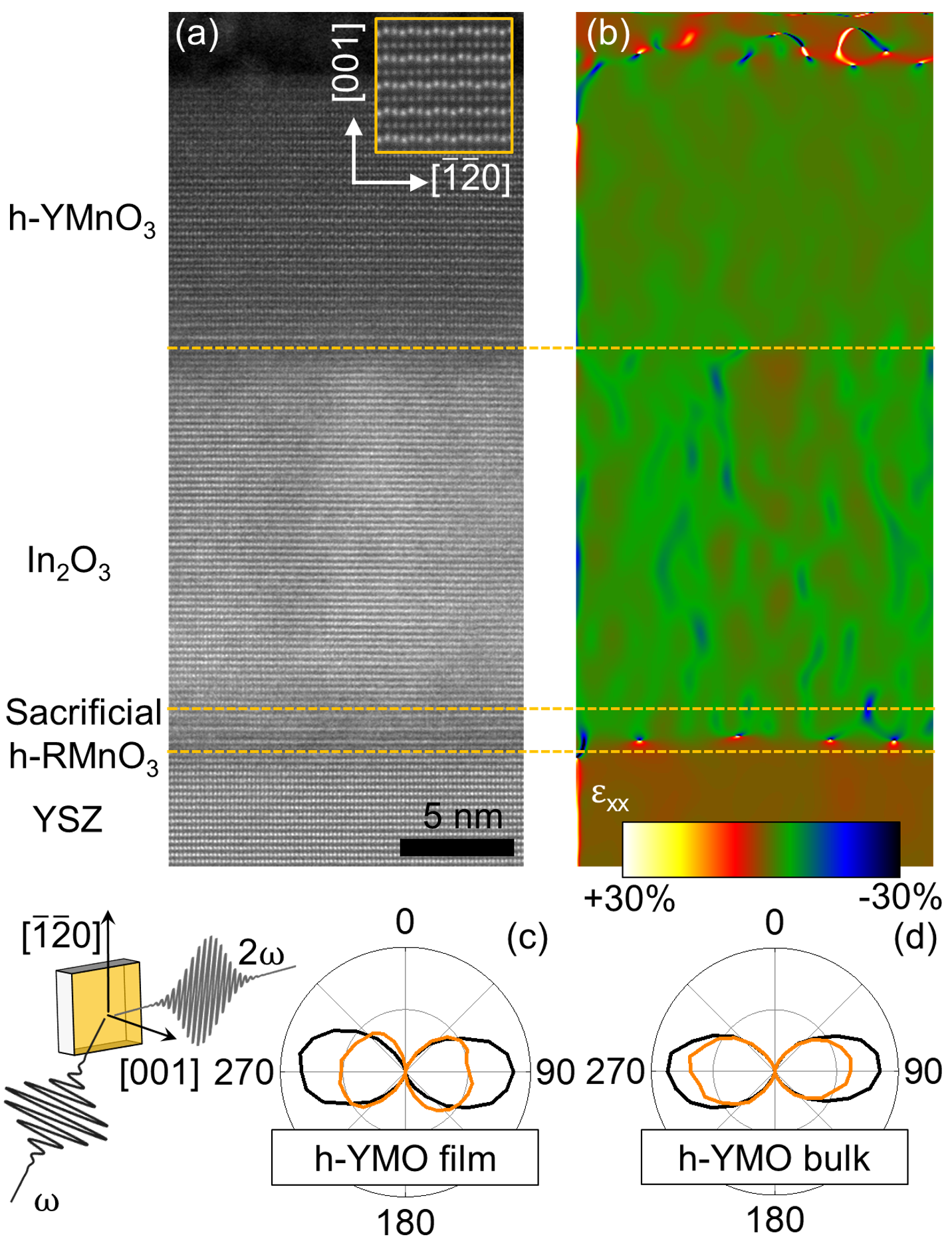}
    \caption{(a) HAADF-STEM image along the [100] zone axis of a 10-unit-cell YMO thin film on RMO-buffered In$_2$O$_3$, here with R = In. Corrugation of the Y plane characteristic of the lattice trimerization driving the ferroelectric order is seen in at high magnification (inset). (b) In-plane strain map by geometric phase analysis of the image in (a). Misfit dislocations are only seen in the sacrificial buffer, such that the lattices of the ITO layer, here shown for isostructural In$_2$O$_3$, and the top YMO layer are decoupled from the YSZ substrate, yet both are lattice matched to the sacrificial layer, in agreement with the XRD analysis [Fig.~\ref{fig:bottomElectrode}(d)]. (c) SHG characterization of an YMO film grown on a pre-annealed YSZ(111) substrate. The SHG polarimetry is measured by rotating the light-polarization of the incident beam ($\lambda = 860$\,nm) from 0\degree\ to 360\degree\ and detecting the component of the SHG light polarized parallel (black) and perpendicular (orange) with respect to the polarization of the incoming light. Here, 0\degree\ and 90\degree\ correspond to light polarized parallel and perpendicular, respectively, to the plane of light reflection. The SHG polarimetry reveals a symmetry compatible with the 6$mm$ point group of YMO in the improper ferroelectric phase with a spontaneous polarization along the normal of the film. (d) Observation of an identical SHG response from a $c$-cut YMO bulk crystal.}
    \label{fig:IFEL}
\end{figure}

We further confirm this structural quality by HAADF-STEM imaging (see Fig.~\ref{fig:IFEL}(a,b)). We find sharp interfaces between the constituent layers, where, indeed, misfit dislocations mediating the lattice mismatch between YSZ and RMO are confined to the sacrificial layer. In contrast, such discontinuities of in-plane lattice parameters are not seen at the top interface between ITO and YMO, where both layers retain the lattice parameter of the sacrificial layer without signatures of thickness-dependent strain relaxation. In addition to this structural quality, the characteristic ferroelectric trimerization of the hexagonal manganites is seen in the film as an up-up-down displacement pattern of Y atoms along the $c$ axis (see inset in Fig.~\ref{fig:IFEL}(a)). We hence note that our growth strategy not only leads to YMO film of significantly improved quality; the demonstration of mutual lattice matching between ferroelectric YMO and conducting ITO films furthermore presents a route to construct coherently strained (metal$|$ improper ferroelectric)-type multilayers combining hexagonal and cubic thin-film oxides.

Finally, we probe the improper ferroelectric polarization in the epitaxial YMO thin films by laser-optical SHG in reflection, see Fig.~\ref{fig:IFEL}(c). SHG denotes a frequency-doubling of light which is especially sensitive to inversion-symmetry breaking in a material, thus relating the SHG signal directly to the presence of the spontaneous polarization as ferroelectric order parameter \cite{Fiebig2005,Denev2011a,Nordlander2018}. In particular, the light-polarization dependence of the SHG response is dictated by the point-group symmetry of the material, which for the ferroelectric phase of YMO is 6$mm$. The emission of SHG light from the YMO thin films indicates that, indeed, these films exhibit a macroscopic polarization. Polarimetry of the SHG signal, shown in Fig.~\ref{fig:IFEL}(c), is compatible with the 6$mm$ point group and further confirms an out-of-plane-oriented polarization in the film [here corresponding to the contribution emitted under $\pm$90\degree\ in Fig.~\ref{fig:IFEL}(c)]. Note that the SHG polarimetry signal is identical to that of a bulk YMO crystal [Fig.~\ref{fig:IFEL}(d)]. In particular, no signature of in-plane polarization components as they would be caused by the aforementioned misoriented nanoinclusions is observed.

\section{Conclusion}

In summary, we have demonstrated layer-by-layer growth of ultrathin YMO thin films on insulating YSZ, with and without insertion of an electrode layer. We find that an ultraflat substrate surface, achieved by YSZ substrate pre-annealing, in combination with a low energy fluence during PLD is key to achieving layer-by-layer growth of epitaxial YMO films with sub-unit-cell thickness control. We have further demonstrated a route to preserve this excellent thin-film quality even after insertion of an intermediate ITO electrode layer. Moreover, with use of a sacrificial RMO layer between the ITO and the substrate, ITO adopts the RMO lattice constant so that strain-relaxed YMO free of misfit dislocations can be grown on top. Hence, we have demonstrated multifold fundamental improvements in the quality of epitaxial hexagonal manganite films. This approach further suggests the feasibility of constructing epitaxial superlattices of metal$|$ferroelectric$|$metal type using the hexagonal manganites as improper room-temperature ferroelectrics. Improper ferroelectrics are presently recognized as a promising class of functional materials because of a variety of properties surpassing those of conventional ferroelectrics. Here our work can advance the implementation of improper ferroelectrics into functional devices considerably.

\acknowledgments
J.N., M.T. and M.F. acknowledge financial support by the EU European Research Council under Advanced Grant Program No.\ 694955-INSEETO. M.D.R. and M.C. acknowledge support by the Swiss National Science Foundation under Project No\ 200021\_175926.


\bibliographystyle{apsrev4-2}
\bibliography{biblio}

\providecommand{\noopsort}[1]{}\providecommand{\singleletter}[1]{#1}%
\begin{thebibliography}{27}%
\makeatletter
\providecommand \@ifxundefined [1]{%
 \@ifx{#1\undefined}
}%
\providecommand \@ifnum [1]{%
 \ifnum #1\expandafter \@firstoftwo
 \else \expandafter \@secondoftwo
 \fi
}%
\providecommand \@ifx [1]{%
 \ifx #1\expandafter \@firstoftwo
 \else \expandafter \@secondoftwo
 \fi
}%
\providecommand \natexlab [1]{#1}%
\providecommand \enquote  [1]{``#1''}%
\providecommand \bibnamefont  [1]{#1}%
\providecommand \bibfnamefont [1]{#1}%
\providecommand \citenamefont [1]{#1}%
\providecommand \href@noop [0]{\@secondoftwo}%
\providecommand \href [0]{\begingroup \@sanitize@url \@href}%
\providecommand \@href[1]{\@@startlink{#1}\@@href}%
\providecommand \@@href[1]{\endgroup#1\@@endlink}%
\providecommand \@sanitize@url [0]{\catcode `\\12\catcode `\$12\catcode
  `\&12\catcode `\#12\catcode `\^12\catcode `\_12\catcode `\%12\relax}%
\providecommand \@@startlink[1]{}%
\providecommand \@@endlink[0]{}%
\providecommand \url  [0]{\begingroup\@sanitize@url \@url }%
\providecommand \@url [1]{\endgroup\@href {#1}{\urlprefix }}%
\providecommand \urlprefix  [0]{URL }%
\providecommand \Eprint [0]{\href }%
\providecommand \doibase [0]{https://doi.org/}%
\providecommand \selectlanguage [0]{\@gobble}%
\providecommand \bibinfo  [0]{\@secondoftwo}%
\providecommand \bibfield  [0]{\@secondoftwo}%
\providecommand \translation [1]{[#1]}%
\providecommand \BibitemOpen [0]{}%
\providecommand \bibitemStop [0]{}%
\providecommand \bibitemNoStop [0]{.\EOS\space}%
\providecommand \EOS [0]{\spacefactor3000\relax}%
\providecommand \BibitemShut  [1]{\csname bibitem#1\endcsname}%
\let\auto@bib@innerbib\@empty
\bibitem [{\citenamefont {Lilienblum}\ \emph {et~al.}(2015)\citenamefont
  {Lilienblum}, \citenamefont {Lottermoser}, \citenamefont {Manz},
  \citenamefont {Selbach}, \citenamefont {Cano},\ and\ \citenamefont
  {Fiebig}}]{Lilienblum2015}%
  \BibitemOpen
  \bibfield  {author} {\bibinfo {author} {\bibfnamefont {M.}~\bibnamefont
  {Lilienblum}}, \bibinfo {author} {\bibfnamefont {T.}~\bibnamefont
  {Lottermoser}}, \bibinfo {author} {\bibfnamefont {S.}~\bibnamefont {Manz}},
  \bibinfo {author} {\bibfnamefont {S.~M.}\ \bibnamefont {Selbach}}, \bibinfo
  {author} {\bibfnamefont {A.}~\bibnamefont {Cano}}, and\ \bibinfo {author}
  {\bibfnamefont {M.}~\bibnamefont {Fiebig}},\ }\href
  {https://doi.org/10.1038/nphys3468} {\bibfield  {journal} {\bibinfo
  {journal} {Nature Physics}\ }\textbf {\bibinfo {volume} {11}},\ \bibinfo
  {pages} {1070} (\bibinfo {year} {2015})}\BibitemShut {NoStop}%
\bibitem [{\citenamefont {Chae}\ \emph {et~al.}(2012)\citenamefont {Chae},
  \citenamefont {Lee}, \citenamefont {Horibe}, \citenamefont {Tanimura},
  \citenamefont {Mori}, \citenamefont {Gao}, \citenamefont {Carr},\ and\
  \citenamefont {Cheong}}]{Chae2012}%
  \BibitemOpen
  \bibfield  {author} {\bibinfo {author} {\bibfnamefont {S.~C.}\ \bibnamefont
  {Chae}}, \bibinfo {author} {\bibfnamefont {N.}~\bibnamefont {Lee}}, \bibinfo
  {author} {\bibfnamefont {Y.}~\bibnamefont {Horibe}}, \bibinfo {author}
  {\bibfnamefont {M.}~\bibnamefont {Tanimura}}, \bibinfo {author}
  {\bibfnamefont {S.}~\bibnamefont {Mori}}, \bibinfo {author} {\bibfnamefont
  {B.}~\bibnamefont {Gao}}, \bibinfo {author} {\bibfnamefont {S.}~\bibnamefont
  {Carr}}, and\ \bibinfo {author} {\bibfnamefont {S.-W.}\ \bibnamefont
  {Cheong}},\ }\href {https://doi.org/10.1103/PhysRevLett.108.167603}
  {\bibfield  {journal} {\bibinfo  {journal} {Physical Review Letters}\
  }\textbf {\bibinfo {volume} {108}},\ \bibinfo {pages} {167603} (\bibinfo
  {year} {2012})}\BibitemShut {NoStop}%
\bibitem [{\citenamefont {Lorenz}(2013)}]{Lorenz2013}%
  \BibitemOpen
  \bibfield  {author} {\bibinfo {author} {\bibfnamefont {B.}~\bibnamefont
  {Lorenz}},\ }\href {https://doi.org/10.1155/2013/497073} {\bibfield
  {journal} {\bibinfo  {journal} {{ISRN} Condensed Matter Physics}\ }\textbf
  {\bibinfo {volume} {2013}},\ \bibinfo {pages} {497073} (\bibinfo {year}
  {2013})}\BibitemShut {NoStop}%
\bibitem [{\citenamefont {Choi}\ \emph {et~al.}(2010)\citenamefont {Choi},
  \citenamefont {Horibe}, \citenamefont {Yi}, \citenamefont {Choi},
  \citenamefont {Wu},\ and\ \citenamefont {Cheong}}]{Choi2010}%
  \BibitemOpen
  \bibfield  {author} {\bibinfo {author} {\bibfnamefont {T.}~\bibnamefont
  {Choi}}, \bibinfo {author} {\bibfnamefont {Y.}~\bibnamefont {Horibe}},
  \bibinfo {author} {\bibfnamefont {H.~T.}\ \bibnamefont {Yi}}, \bibinfo
  {author} {\bibfnamefont {Y.~J.}\ \bibnamefont {Choi}}, \bibinfo {author}
  {\bibfnamefont {W.}~\bibnamefont {Wu}}, and\ \bibinfo {author} {\bibfnamefont
  {S.-W.}\ \bibnamefont {Cheong}},\ }\href {https://doi.org/10.1038/nmat2632}
  {\bibfield  {journal} {\bibinfo  {journal} {Nature Materials}\ }\textbf
  {\bibinfo {volume} {9}},\ \bibinfo {pages} {253–258} (\bibinfo {year}
  {2010})}\BibitemShut {NoStop}%
\bibitem [{\citenamefont {Jungk}\ \emph {et~al.}(2010)\citenamefont {Jungk},
  \citenamefont {Hoffmann}, \citenamefont {Fiebig},\ and\ \citenamefont
  {Soergel}}]{Jungk2010}%
  \BibitemOpen
  \bibfield  {author} {\bibinfo {author} {\bibfnamefont {T.}~\bibnamefont
  {Jungk}}, \bibinfo {author} {\bibfnamefont {{\'{A}}.}~\bibnamefont
  {Hoffmann}}, \bibinfo {author} {\bibfnamefont {M.}~\bibnamefont {Fiebig}},
  and\ \bibinfo {author} {\bibfnamefont {E.}~\bibnamefont {Soergel}},\ }\href
  {https://doi.org/10.1063/1.3460286} {\bibfield  {journal} {\bibinfo
  {journal} {Applied Physics Letters}\ }\textbf {\bibinfo {volume} {97}},\
  \bibinfo {pages} {12904} (\bibinfo {year} {2010})}\BibitemShut {NoStop}%
\bibitem [{\citenamefont {Meier}\ \emph {et~al.}(2012)\citenamefont {Meier},
  \citenamefont {Seidel}, \citenamefont {Cano}, \citenamefont {Delaney},
  \citenamefont {Kumagai}, \citenamefont {Mostovoy}, \citenamefont {Spaldin},
  \citenamefont {Ramesh},\ and\ \citenamefont {Fiebig}}]{Meier2012a}%
  \BibitemOpen
  \bibfield  {author} {\bibinfo {author} {\bibfnamefont {D.}~\bibnamefont
  {Meier}}, \bibinfo {author} {\bibfnamefont {J.}~\bibnamefont {Seidel}},
  \bibinfo {author} {\bibfnamefont {A.}~\bibnamefont {Cano}}, \bibinfo {author}
  {\bibfnamefont {K.}~\bibnamefont {Delaney}}, \bibinfo {author} {\bibfnamefont
  {Y.}~\bibnamefont {Kumagai}}, \bibinfo {author} {\bibfnamefont
  {M.}~\bibnamefont {Mostovoy}}, \bibinfo {author} {\bibfnamefont {N.~A.}\
  \bibnamefont {Spaldin}}, \bibinfo {author} {\bibfnamefont {R.}~\bibnamefont
  {Ramesh}}, and\ \bibinfo {author} {\bibfnamefont {M.}~\bibnamefont
  {Fiebig}},\ }\href {http://dx.doi.org/10.1038/nmat3249
  http://10.0.4.14/nmat3249
  https://www.nature.com/articles/nmat3249{\#}supplementary-information}
  {\bibfield  {journal} {\bibinfo  {journal} {Nature Materials}\ }\textbf
  {\bibinfo {volume} {11}},\ \bibinfo {pages} {284} (\bibinfo {year}
  {2012})}\BibitemShut {NoStop}%
\bibitem [{\citenamefont {Fiebig}\ \emph {et~al.}(2002)\citenamefont {Fiebig},
  \citenamefont {Lottermoser}, \citenamefont {Fr{\"{o}}hlich}, \citenamefont
  {Goltsev},\ and\ \citenamefont {Pisarev}}]{Fiebig2002}%
  \BibitemOpen
  \bibfield  {author} {\bibinfo {author} {\bibfnamefont {M.}~\bibnamefont
  {Fiebig}}, \bibinfo {author} {\bibfnamefont {T.}~\bibnamefont {Lottermoser}},
  \bibinfo {author} {\bibfnamefont {D.}~\bibnamefont {Fr{\"{o}}hlich}},
  \bibinfo {author} {\bibfnamefont {A.~V.}\ \bibnamefont {Goltsev}}, and\
  \bibinfo {author} {\bibfnamefont {R.~V.}\ \bibnamefont {Pisarev}},\ }\href
  {https://doi.org/10.1038/nature01077} {\bibfield  {journal} {\bibinfo
  {journal} {Nature}\ }\textbf {\bibinfo {volume} {419}},\ \bibinfo {pages}
  {818} (\bibinfo {year} {2002})}\BibitemShut {NoStop}%
\bibitem [{\citenamefont {Nordlander}\ \emph {et~al.}(2019)\citenamefont
  {Nordlander}, \citenamefont {Campanini}, \citenamefont {Rossell},
  \citenamefont {Erni}, \citenamefont {Meier}, \citenamefont {Cano},
  \citenamefont {Spaldin}, \citenamefont {Fiebig},\ and\ \citenamefont
  {Trassin}}]{Nordlander2019}%
  \BibitemOpen
  \bibfield  {author} {\bibinfo {author} {\bibfnamefont {J.}~\bibnamefont
  {Nordlander}}, \bibinfo {author} {\bibfnamefont {M.}~\bibnamefont
  {Campanini}}, \bibinfo {author} {\bibfnamefont {M.~D.}\ \bibnamefont
  {Rossell}}, \bibinfo {author} {\bibfnamefont {R.}~\bibnamefont {Erni}},
  \bibinfo {author} {\bibfnamefont {Q.~N.}\ \bibnamefont {Meier}}, \bibinfo
  {author} {\bibfnamefont {A.}~\bibnamefont {Cano}}, \bibinfo {author}
  {\bibfnamefont {N.~A.}\ \bibnamefont {Spaldin}}, \bibinfo {author}
  {\bibfnamefont {M.}~\bibnamefont {Fiebig}}, and\ \bibinfo {author}
  {\bibfnamefont {M.}~\bibnamefont {Trassin}},\ }\href
  {https://doi.org/10.1038/s41467-019-13474-x} {\bibfield  {journal} {\bibinfo
  {journal} {Nature Communications}\ }\textbf {\bibinfo {volume} {10}},\
  \bibinfo {pages} {5591} (\bibinfo {year} {2019})}\BibitemShut {NoStop}%
\bibitem [{\citenamefont {Fujimura}\ \emph {et~al.}(1996)\citenamefont
  {Fujimura}, \citenamefont {Azuma}, \citenamefont {Aoki}, \citenamefont
  {Yoshimura},\ and\ \citenamefont {Ito}}]{Fujimura1996a}%
  \BibitemOpen
  \bibfield  {author} {\bibinfo {author} {\bibfnamefont {N.}~\bibnamefont
  {Fujimura}}, \bibinfo {author} {\bibfnamefont {S.-i.}\ \bibnamefont {Azuma}},
  \bibinfo {author} {\bibfnamefont {N.}~\bibnamefont {Aoki}}, \bibinfo {author}
  {\bibfnamefont {T.}~\bibnamefont {Yoshimura}}, and\ \bibinfo {author}
  {\bibfnamefont {T.}~\bibnamefont {Ito}},\ }\href
  {https://doi.org/10.1063/1.363719} {\bibfield  {journal} {\bibinfo  {journal}
  {Journal of Applied Physics}\ }\textbf {\bibinfo {volume} {80}},\ \bibinfo
  {pages} {7084} (\bibinfo {year} {1996})}\BibitemShut {NoStop}%
\bibitem [{\citenamefont {Dho}\ \emph {et~al.}(2004)\citenamefont {Dho},
  \citenamefont {Leung}, \citenamefont {MacManus-Driscoll},\ and\ \citenamefont
  {Blamire}}]{Dho2004}%
  \BibitemOpen
  \bibfield  {author} {\bibinfo {author} {\bibfnamefont {J.}~\bibnamefont
  {Dho}}, \bibinfo {author} {\bibfnamefont {C.~W.}\ \bibnamefont {Leung}},
  \bibinfo {author} {\bibfnamefont {J.~L.}\ \bibnamefont {MacManus-Driscoll}},
  and\ \bibinfo {author} {\bibfnamefont {M.~G.}\ \bibnamefont {Blamire}},\
  }\href {https://doi.org/10.1016/j.jcrysgro.2004.04.028} {\bibfield  {journal}
  {\bibinfo  {journal} {Journal of Crystal Growth}\ }\textbf {\bibinfo {volume}
  {267}},\ \bibinfo {pages} {548} (\bibinfo {year} {2004})}\BibitemShut
  {NoStop}%
\bibitem [{\citenamefont {Mart{\'{i}}}\ \emph {et~al.}(2007)\citenamefont
  {Mart{\'{i}}}, \citenamefont {S{\'{a}}nchez}, \citenamefont {Hrabovsky},
  \citenamefont {Fontcuberta}, \citenamefont {Laukhin}, \citenamefont
  {Skumryev}, \citenamefont {Garc{\'{i}}a-Cuenca}, \citenamefont {Ferrater},
  \citenamefont {Varela}, \citenamefont {L{\"{u}}ders}, \citenamefont {Bobo},
  \citenamefont {Estrad{\'{e}}}, \citenamefont {Arbiol},\ and\ \citenamefont
  {Peir{\'{o}}}}]{Marti2007}%
  \BibitemOpen
  \bibfield  {author} {\bibinfo {author} {\bibfnamefont {X.}~\bibnamefont
  {Mart{\'{i}}}}, \bibinfo {author} {\bibfnamefont {F.}~\bibnamefont
  {S{\'{a}}nchez}}, \bibinfo {author} {\bibfnamefont {D.}~\bibnamefont
  {Hrabovsky}}, \bibinfo {author} {\bibfnamefont {J.}~\bibnamefont
  {Fontcuberta}}, \bibinfo {author} {\bibfnamefont {V.}~\bibnamefont
  {Laukhin}}, \bibinfo {author} {\bibfnamefont {V.}~\bibnamefont {Skumryev}},
  \bibinfo {author} {\bibfnamefont {M.~V.}\ \bibnamefont
  {Garc{\'{i}}a-Cuenca}}, \bibinfo {author} {\bibfnamefont {C.}~\bibnamefont
  {Ferrater}}, \bibinfo {author} {\bibfnamefont {M.}~\bibnamefont {Varela}},
  \bibinfo {author} {\bibfnamefont {U.}~\bibnamefont {L{\"{u}}ders}}, \bibinfo
  {author} {\bibfnamefont {J.~F.}\ \bibnamefont {Bobo}}, \bibinfo {author}
  {\bibfnamefont {S.}~\bibnamefont {Estrad{\'{e}}}}, \bibinfo {author}
  {\bibfnamefont {J.}~\bibnamefont {Arbiol}}, and\ \bibinfo {author}
  {\bibfnamefont {F.}~\bibnamefont {Peir{\'{o}}}},\ }\href
  {https://doi.org/10.1016/j.jcrysgro.2006.11.272} {\bibfield  {journal}
  {\bibinfo  {journal} {Journal of Crystal Growth}\ }\textbf {\bibinfo {volume}
  {299}},\ \bibinfo {pages} {288} (\bibinfo {year} {2007})}\BibitemShut
  {NoStop}%
\bibitem [{\citenamefont {Imada}\ \emph {et~al.}(1998)\citenamefont {Imada},
  \citenamefont {Shouriki}, \citenamefont {Tokumitsu},\ and\ \citenamefont
  {Ishiwara}}]{Imada1998}%
  \BibitemOpen
  \bibfield  {author} {\bibinfo {author} {\bibfnamefont {S.}~\bibnamefont
  {Imada}}, \bibinfo {author} {\bibfnamefont {S.}~\bibnamefont {Shouriki}},
  \bibinfo {author} {\bibfnamefont {E.}~\bibnamefont {Tokumitsu}}, and\
  \bibinfo {author} {\bibfnamefont {H.}~\bibnamefont {Ishiwara}},\ }\href
  {https://doi.org/10.1143/JJAP.37.6497} {\bibfield  {journal} {\bibinfo
  {journal} {Japanese Journal of Applied Physics}\ }\textbf {\bibinfo {volume}
  {37}},\ \bibinfo {pages} {6497} (\bibinfo {year} {1998})}\BibitemShut
  {NoStop}%
\bibitem [{\citenamefont {Cheng}\ \emph {et~al.}(2018)\citenamefont {Cheng},
  \citenamefont {Xu}, \citenamefont {Deng}, \citenamefont {Han}, \citenamefont
  {Bao}, \citenamefont {Ma}, \citenamefont {Nan}, \citenamefont {Duan},
  \citenamefont {Bellaiche}, \citenamefont {Zhu},\ and\ \citenamefont
  {Zhu}}]{Cheng2018}%
  \BibitemOpen
  \bibfield  {author} {\bibinfo {author} {\bibfnamefont {S.}~\bibnamefont
  {Cheng}}, \bibinfo {author} {\bibfnamefont {C.}~\bibnamefont {Xu}}, \bibinfo
  {author} {\bibfnamefont {S.}~\bibnamefont {Deng}}, \bibinfo {author}
  {\bibfnamefont {M.-G.}\ \bibnamefont {Han}}, \bibinfo {author} {\bibfnamefont
  {S.}~\bibnamefont {Bao}}, \bibinfo {author} {\bibfnamefont {J.}~\bibnamefont
  {Ma}}, \bibinfo {author} {\bibfnamefont {C.}~\bibnamefont {Nan}}, \bibinfo
  {author} {\bibfnamefont {W.}~\bibnamefont {Duan}}, \bibinfo {author}
  {\bibfnamefont {L.}~\bibnamefont {Bellaiche}}, \bibinfo {author}
  {\bibfnamefont {Y.}~\bibnamefont {Zhu}}, and\ \bibinfo {author}
  {\bibfnamefont {J.}~\bibnamefont {Zhu}},\ }\href
  {https://doi.org/10.1126/sciadv.aar4298} {\bibfield  {journal} {\bibinfo
  {journal} {Science Advances}\ }\textbf {\bibinfo {volume} {4}},\ \bibinfo
  {pages} {eaar4298} (\bibinfo {year} {2018})}\BibitemShut {NoStop}%
\bibitem [{\citenamefont {Imada}\ \emph {et~al.}(2001)\citenamefont {Imada},
  \citenamefont {Kuraoka}, \citenamefont {Tokumitsu},\ and\ \citenamefont
  {Ishiwara}}]{Imada2001}%
  \BibitemOpen
  \bibfield  {author} {\bibinfo {author} {\bibfnamefont {S.}~\bibnamefont
  {Imada}}, \bibinfo {author} {\bibfnamefont {T.}~\bibnamefont {Kuraoka}},
  \bibinfo {author} {\bibfnamefont {E.}~\bibnamefont {Tokumitsu}}, and\
  \bibinfo {author} {\bibfnamefont {H.}~\bibnamefont {Ishiwara}},\ }\href
  {https://doi.org/10.1143/JJAP.40.666} {\bibfield  {journal} {\bibinfo
  {journal} {Japanese Journal of Applied Physics}\ }\textbf {\bibinfo {volume}
  {40}},\ \bibinfo {pages} {666} (\bibinfo {year} {2001})}\BibitemShut
  {NoStop}%
\bibitem [{\citenamefont {Wu}\ \emph {et~al.}(2011)\citenamefont {Wu},
  \citenamefont {Chen}, \citenamefont {Chen}, \citenamefont {Hsieh},
  \citenamefont {Luo}, \citenamefont {Uen}, \citenamefont {Juang},
  \citenamefont {Lin}, \citenamefont {Kobayashi},\ and\ \citenamefont
  {Gospodinov}}]{Wu2011}%
  \BibitemOpen
  \bibfield  {author} {\bibinfo {author} {\bibfnamefont {K.~H.}\ \bibnamefont
  {Wu}}, \bibinfo {author} {\bibfnamefont {H.-J.}\ \bibnamefont {Chen}},
  \bibinfo {author} {\bibfnamefont {Y.~T.}\ \bibnamefont {Chen}}, \bibinfo
  {author} {\bibfnamefont {C.~C.}\ \bibnamefont {Hsieh}}, \bibinfo {author}
  {\bibfnamefont {C.~W.}\ \bibnamefont {Luo}}, \bibinfo {author} {\bibfnamefont
  {T.~M.}\ \bibnamefont {Uen}}, \bibinfo {author} {\bibfnamefont {J.~Y.}\
  \bibnamefont {Juang}}, \bibinfo {author} {\bibfnamefont {J.-Y.}\ \bibnamefont
  {Lin}}, \bibinfo {author} {\bibfnamefont {T.}~\bibnamefont {Kobayashi}}, and\
  \bibinfo {author} {\bibfnamefont {M.}~\bibnamefont {Gospodinov}},\ }\href
  {https://doi.org/10.1209/0295-5075/94/27006} {\bibfield  {journal} {\bibinfo
  {journal} {EPL (Europhysics Letters)}\ }\textbf {\bibinfo {volume} {94}},\
  \bibinfo {pages} {27006} (\bibinfo {year} {2011})}\BibitemShut {NoStop}%
\bibitem [{\citenamefont {Dubourdieu}\ \emph {et~al.}(2007)\citenamefont
  {Dubourdieu}, \citenamefont {Huot}, \citenamefont {Gelard}, \citenamefont
  {Roussel}, \citenamefont {Lebedev},\ and\ \citenamefont {{Van
  Tendeloo}}}]{Dubourdieu2007}%
  \BibitemOpen
  \bibfield  {author} {\bibinfo {author} {\bibfnamefont {C.}~\bibnamefont
  {Dubourdieu}}, \bibinfo {author} {\bibfnamefont {G.}~\bibnamefont {Huot}},
  \bibinfo {author} {\bibfnamefont {I.}~\bibnamefont {Gelard}}, \bibinfo
  {author} {\bibfnamefont {H.}~\bibnamefont {Roussel}}, \bibinfo {author}
  {\bibfnamefont {O.}~\bibnamefont {Lebedev}}, and\ \bibinfo {author}
  {\bibfnamefont {G.}~\bibnamefont {{Van Tendeloo}}},\ }\href
  {https://doi.org/10.1080/09500830601137173} {\bibfield  {journal} {\bibinfo
  {journal} {Philosophical Magazine Letters}\ }\textbf {\bibinfo {volume}
  {87}},\ \bibinfo {pages} {203} (\bibinfo {year} {2007})}\BibitemShut
  {NoStop}%
\bibitem [{\citenamefont {Laukhin}\ \emph {et~al.}(2006)\citenamefont
  {Laukhin}, \citenamefont {Skumryev}, \citenamefont {Mart{\'{i}}},
  \citenamefont {Hrabovsky}, \citenamefont {S{\'{a}}nchez}, \citenamefont
  {Garc{\'{i}}a-Cuenca}, \citenamefont {Ferrater}, \citenamefont {Varela},
  \citenamefont {L{\"{u}}ders}, \citenamefont {Bobo},\ and\ \citenamefont
  {Fontcuberta}}]{Laukhin2006}%
  \BibitemOpen
  \bibfield  {author} {\bibinfo {author} {\bibfnamefont {V.}~\bibnamefont
  {Laukhin}}, \bibinfo {author} {\bibfnamefont {V.}~\bibnamefont {Skumryev}},
  \bibinfo {author} {\bibfnamefont {X.}~\bibnamefont {Mart{\'{i}}}}, \bibinfo
  {author} {\bibfnamefont {D.}~\bibnamefont {Hrabovsky}}, \bibinfo {author}
  {\bibfnamefont {F.}~\bibnamefont {S{\'{a}}nchez}}, \bibinfo {author}
  {\bibfnamefont {M.~V.}\ \bibnamefont {Garc{\'{i}}a-Cuenca}}, \bibinfo
  {author} {\bibfnamefont {C.}~\bibnamefont {Ferrater}}, \bibinfo {author}
  {\bibfnamefont {M.}~\bibnamefont {Varela}}, \bibinfo {author} {\bibfnamefont
  {U.}~\bibnamefont {L{\"{u}}ders}}, \bibinfo {author} {\bibfnamefont {J.~F.}\
  \bibnamefont {Bobo}}, and\ \bibinfo {author} {\bibfnamefont {J.}~\bibnamefont
  {Fontcuberta}},\ }\href {https://doi.org/10.1103/PhysRevLett.97.227201}
  {\bibfield  {journal} {\bibinfo  {journal} {Physical Review Letters}\
  }\textbf {\bibinfo {volume} {97}},\ \bibinfo {pages} {227201} (\bibinfo
  {year} {2006})}\BibitemShut {NoStop}%
\bibitem [{FRW()}]{FRWRtools}%
  \BibitemOpen
  \href@noop {} {\bibinfo {title} {{FRWRtools plugin}}},\ \bibinfo
  {howpublished}
  {\url{https://www.physics.hu-berlin.de/en/sem/software/software_frwrtools}},\
  \bibinfo {note} {accessed: June, 2020}\BibitemShut {NoStop}%
\bibitem [{\citenamefont {Kordel}\ \emph {et~al.}(2009)\citenamefont {Kordel},
  \citenamefont {Wehrenfennig}, \citenamefont {Meier}, \citenamefont
  {Lottermoser}, \citenamefont {Fiebig}, \citenamefont {G{\'{e}}lard},
  \citenamefont {Dubourdieu}, \citenamefont {Kim}, \citenamefont {Schultz},\
  and\ \citenamefont {D{\"{o}}rr}}]{Kordel2009}%
  \BibitemOpen
  \bibfield  {author} {\bibinfo {author} {\bibfnamefont {T.}~\bibnamefont
  {Kordel}}, \bibinfo {author} {\bibfnamefont {C.}~\bibnamefont
  {Wehrenfennig}}, \bibinfo {author} {\bibfnamefont {D.}~\bibnamefont {Meier}},
  \bibinfo {author} {\bibfnamefont {T.}~\bibnamefont {Lottermoser}}, \bibinfo
  {author} {\bibfnamefont {M.}~\bibnamefont {Fiebig}}, \bibinfo {author}
  {\bibfnamefont {I.}~\bibnamefont {G{\'{e}}lard}}, \bibinfo {author}
  {\bibfnamefont {C.}~\bibnamefont {Dubourdieu}}, \bibinfo {author}
  {\bibfnamefont {J.-W.}\ \bibnamefont {Kim}}, \bibinfo {author} {\bibfnamefont
  {L.}~\bibnamefont {Schultz}}, and\ \bibinfo {author} {\bibfnamefont
  {K.}~\bibnamefont {D{\"{o}}rr}},\ }\href
  {https://doi.org/10.1103/PhysRevB.80.045409} {\bibfield  {journal} {\bibinfo
  {journal} {Physical Review B}\ }\textbf {\bibinfo {volume} {80}},\ \bibinfo
  {pages} {045409} (\bibinfo {year} {2009})}\BibitemShut {NoStop}%
\bibitem [{\citenamefont {Jehanathan}\ \emph {et~al.}(2010)\citenamefont
  {Jehanathan}, \citenamefont {Lebedev}, \citenamefont {G{\'{e}}lard},
  \citenamefont {Dubourdieu},\ and\ \citenamefont {{Van
  Tendeloo}}}]{Jehanathan2010}%
  \BibitemOpen
  \bibfield  {author} {\bibinfo {author} {\bibfnamefont {N.}~\bibnamefont
  {Jehanathan}}, \bibinfo {author} {\bibfnamefont {O.}~\bibnamefont {Lebedev}},
  \bibinfo {author} {\bibfnamefont {I.}~\bibnamefont {G{\'{e}}lard}}, \bibinfo
  {author} {\bibfnamefont {C.}~\bibnamefont {Dubourdieu}}, and\ \bibinfo
  {author} {\bibfnamefont {G.}~\bibnamefont {{Van Tendeloo}}},\ }\href
  {https://doi.org/10.1088/0957-4484/21/7/075705} {\bibfield  {journal}
  {\bibinfo  {journal} {Nanotechnology}\ }\textbf {\bibinfo {volume} {21}},\
  \bibinfo {pages} {075705} (\bibinfo {year} {2010})}\BibitemShut {NoStop}%
\bibitem [{\citenamefont {Ohring}(1992)}]{Ohring1992}%
  \BibitemOpen
  \bibfield  {author} {\bibinfo {author} {\bibfnamefont {M.}~\bibnamefont
  {Ohring}},\ }\href@noop {} {\emph {\bibinfo {title} {The materials science of
  thin films}}}\ (\bibinfo  {publisher} {Academic Press},\ \bibinfo {address}
  {Boston},\ \bibinfo {year} {1992})\BibitemShut {NoStop}%
\bibitem [{\citenamefont {Honke}\ \emph {et~al.}(2004)\citenamefont {Honke},
  \citenamefont {Fujioka}, \citenamefont {Ohta},\ and\ \citenamefont
  {Oshima}}]{Honke2004}%
  \BibitemOpen
  \bibfield  {author} {\bibinfo {author} {\bibfnamefont {T.}~\bibnamefont
  {Honke}}, \bibinfo {author} {\bibfnamefont {H.}~\bibnamefont {Fujioka}},
  \bibinfo {author} {\bibfnamefont {J.}~\bibnamefont {Ohta}}, and\ \bibinfo
  {author} {\bibfnamefont {M.}~\bibnamefont {Oshima}},\ }\href
  {https://doi.org/10.1116/1.1809127} {\bibfield  {journal} {\bibinfo
  {journal} {Journal of Vacuum Science {\&} Technology A: Vacuum, Surfaces, and
  Films}\ }\textbf {\bibinfo {volume} {22}},\ \bibinfo {pages} {2487} (\bibinfo
  {year} {2004})}\BibitemShut {NoStop}%
\bibitem [{\citenamefont {Ohta}\ \emph {et~al.}(2000)\citenamefont {Ohta},
  \citenamefont {Orita}, \citenamefont {Hirano}, \citenamefont {Tanji},
  \citenamefont {Kawazoe},\ and\ \citenamefont {Hosono}}]{Ohta2000}%
  \BibitemOpen
  \bibfield  {author} {\bibinfo {author} {\bibfnamefont {H.}~\bibnamefont
  {Ohta}}, \bibinfo {author} {\bibfnamefont {M.}~\bibnamefont {Orita}},
  \bibinfo {author} {\bibfnamefont {M.}~\bibnamefont {Hirano}}, \bibinfo
  {author} {\bibfnamefont {H.}~\bibnamefont {Tanji}}, \bibinfo {author}
  {\bibfnamefont {H.}~\bibnamefont {Kawazoe}}, and\ \bibinfo {author}
  {\bibfnamefont {H.}~\bibnamefont {Hosono}},\ }\href
  {https://doi.org/10.1063/1.126461} {\bibfield  {journal} {\bibinfo  {journal}
  {Applied Physics Letters}\ }\textbf {\bibinfo {volume} {76}},\ \bibinfo
  {pages} {2740} (\bibinfo {year} {2000})}\BibitemShut {NoStop}%
\bibitem [{\citenamefont {Kim}\ \emph {et~al.}(2002)\citenamefont {Kim},
  \citenamefont {Horwitz}, \citenamefont {Kim}, \citenamefont {Kafafi},\ and\
  \citenamefont {Chrisey}}]{Kim2002}%
  \BibitemOpen
  \bibfield  {author} {\bibinfo {author} {\bibfnamefont {H.}~\bibnamefont
  {Kim}}, \bibinfo {author} {\bibfnamefont {J.~S.}\ \bibnamefont {Horwitz}},
  \bibinfo {author} {\bibfnamefont {W.~H.}\ \bibnamefont {Kim}}, \bibinfo
  {author} {\bibfnamefont {Z.~H.}\ \bibnamefont {Kafafi}}, and\ \bibinfo
  {author} {\bibfnamefont {D.~B.}\ \bibnamefont {Chrisey}},\ }\href
  {https://doi.org/10.1063/1.1461068} {\bibfield  {journal} {\bibinfo
  {journal} {Journal of Applied Physics}\ }\textbf {\bibinfo {volume} {91}},\
  \bibinfo {pages} {5371} (\bibinfo {year} {2002})}\BibitemShut {NoStop}%
\bibitem [{\citenamefont {Fiebig}\ \emph {et~al.}(2005)\citenamefont {Fiebig},
  \citenamefont {Pavlov},\ and\ \citenamefont {Pisarev}}]{Fiebig2005}%
  \BibitemOpen
  \bibfield  {author} {\bibinfo {author} {\bibfnamefont {M.}~\bibnamefont
  {Fiebig}}, \bibinfo {author} {\bibfnamefont {V.~V.}\ \bibnamefont {Pavlov}},
  and\ \bibinfo {author} {\bibfnamefont {R.~V.}\ \bibnamefont {Pisarev}},\
  }\href {https://doi.org/10.1364/JOSAB.22.000096} {\bibfield  {journal}
  {\bibinfo  {journal} {Journal of the Optical Society of America B}\ }\textbf
  {\bibinfo {volume} {22}},\ \bibinfo {pages} {96} (\bibinfo {year}
  {2005})}\BibitemShut {NoStop}%
\bibitem [{\citenamefont {Denev}\ \emph {et~al.}(2011)\citenamefont {Denev},
  \citenamefont {Lummen}, \citenamefont {Barnes}, \citenamefont {Kumar},\ and\
  \citenamefont {Gopalan}}]{Denev2011a}%
  \BibitemOpen
  \bibfield  {author} {\bibinfo {author} {\bibfnamefont {S.~A.}\ \bibnamefont
  {Denev}}, \bibinfo {author} {\bibfnamefont {T.~T.~A.}\ \bibnamefont
  {Lummen}}, \bibinfo {author} {\bibfnamefont {E.}~\bibnamefont {Barnes}},
  \bibinfo {author} {\bibfnamefont {A.}~\bibnamefont {Kumar}}, and\ \bibinfo
  {author} {\bibfnamefont {V.}~\bibnamefont {Gopalan}},\ }\href
  {https://doi.org/10.1111/j.1551-2916.2011.04740.x} {\bibfield  {journal}
  {\bibinfo  {journal} {Journal of the American Ceramic Society}\ }\textbf
  {\bibinfo {volume} {94}},\ \bibinfo {pages} {2699} (\bibinfo {year}
  {2011})}\BibitemShut {NoStop}%
\bibitem [{\citenamefont {Nordlander}\ \emph {et~al.}(2018)\citenamefont
  {Nordlander}, \citenamefont {{De Luca}}, \citenamefont {Strkalj},
  \citenamefont {Fiebig},\ and\ \citenamefont {Trassin}}]{Nordlander2018}%
  \BibitemOpen
  \bibfield  {author} {\bibinfo {author} {\bibfnamefont {J.}~\bibnamefont
  {Nordlander}}, \bibinfo {author} {\bibfnamefont {G.}~\bibnamefont {{De
  Luca}}}, \bibinfo {author} {\bibfnamefont {N.}~\bibnamefont {Strkalj}},
  \bibinfo {author} {\bibfnamefont {M.}~\bibnamefont {Fiebig}}, and\ \bibinfo
  {author} {\bibfnamefont {M.}~\bibnamefont {Trassin}},\ }\href
  {https://doi.org/10.3390/app8040570} {\bibfield  {journal} {\bibinfo
  {journal} {Applied Sciences}\ }\textbf {\bibinfo {volume} {8}},\ \bibinfo
  {pages} {570} (\bibinfo {year} {2018})}\BibitemShut {NoStop}%
\end{thebibliography}%


\providecommand{\noopsort}[1]{}\providecommand{\singleletter}[1]{#1}%
%

\end{document}